# A Sub-Parsec Origin for Rapid Accretion-State Transitions in Radio-Loud AGN

David Garofalo

Department of Physics, Kennesaw State University, USA

Abstract

I investigate the physical origin and timescale of accretion-state transitions in radio-loud active galactic nuclei, focusing on the evolution from FRII high-excitation radio galaxies (HERGs) to FRII low-excitation radio galaxies (LERGs) within the framework of Garofalo, Evans & Sambruna (2010). I show that the required transition timescale of $\sim 4 \times 10^6$yr arises naturally if the accretion rate is regulated at sub-parsec radii ($\sim 0.1 - 0.3$ pc), corresponding to viscous evolution at $\sim 10^3 - 10^4\ R_g$for black hole masses $10^8 - 10^9\ M_\odot$. These radii coincide with the broad-line region and inner dusty torus, where multiphase gas dynamics regulate the effective inward mass flux on Myr timescales. I argue that powerful FRII jet feedback heats and perturbs the surrounding medium, reducing the effective mass supply reaching the inner disk. Once the accretion rate falls below the critical threshold $\dot{m}_{\rm crit} \sim 0.01$, the inner flow rapidly transitions to an advection-dominated accretion flow (ADAF). Observations of molecular gas reservoirs in radio galaxies are compatible with this picture because the transition requires suppression of inward fueling rather than complete removal of cold gas. I conclude that rapid, internally driven evolution provides a physically consistent basis for the FRII HERG →FRII LERG transition while linking radio morphology, accretion state, and cosmic evolution.

## 1. Introduction

Radio-loud active galactic nuclei (AGN) are commonly divided into high-excitation radio galaxies (HERGs), associated with radiatively efficient accretion, and low-excitation radio galaxies (LERGs), associated with radiatively inefficient flows. Within the framework of Garofalo, Evans & Sambruna (2010), powerful FRII HERGs evolve into FRII LERGs through a transition from a thin disk to an ADAF while preserving large-scale jet morphology. A central question is whether this transition can occur rapidly enough to remain consistent with the counter-rotating FRII phase.

While Garofalo, Evans & Sambruna (2010) proposed that powerful FRII HERGs evolve into FRII LERGs through jet-driven suppression of accretion, no explicit calculation of the physical scale or viscous timescale associated with the required accretion-rate decline was presented. The purpose of the present Research Note is to show that the observed few-Myr transition timescale arises naturally from viscous evolution at sub-parsec fueling radii.

I show that a transition timescale of $\sim 4 \times 10^6$yr arises naturally if the accretion rate is regulated at sub-parsec radii ($\sim 0.1 - 0.3$ pc or $\sim 10^3 - 10^4\ R_g$ where $R_g \equiv GM/c^2$) for black hole masses $10^8 - 10^9\ M_\odot$. These radii correspond to the broad-line region and inner dusty torus, where multiphase gas dynamics and inefficient angular momentum transport regulate the effective inward mass flux on Myr timescales. Once the accretion rate falls below $\dot{m}_{\rm crit} \sim 0.01$, the inner flow rapidly transitions to an ADAF on much shorter viscous timescales ($\lesssim 10^5$ yr).

In this framework, FRII jet feedback heats and perturbs the surrounding medium, reducing the effective mass supply reaching the nucleus without requiring complete removal of the cold gas reservoir. I argue that this provides a physically consistent basis for the FRII HERG →FRII LERG transition while naturally linking radio morphology, accretion state, and cosmic evolution.

2. Jet-Driven Suppression of Accretion and the Transition Timescale

The transition from a radiatively efficient thin disk to a radiatively inefficient flow occurs when the accretion rate falls below the critical threshold $\dot{m}_{\rm crit} \sim 10^{-2}$ (Narayan & Yi 1995; Yuan & Narayan 2014). A central question is whether the accretion rate in powerful FRII HERGs can decline to this level on Myr timescales, as required in the framework of Garofalo, Evans & Sambruna (2010).

In this picture, the transition from FRII HERG to FRII LERG is driven by feedback from the FRII jet itself. Powerful jets heat and rarefy the surrounding interstellar medium, reducing the efficiency of inward gas transport. For characteristic jet powers

$$P_{\rm jet} \sim 10^{46} \text{ erg/s}$$

(e.g. Best & Heckman 2014) sustained over $\sim 4 \times 10^6$ yr, the total injected energy reaches $\sim 10^{60}$erg, exceeding the binding energy of the central cold gas reservoir. Only modest coupling is therefore required to substantially suppress the inflow. Importantly, the model does not require complete removal of the cold gas reservoir. ALMA observations reveal substantial molecular gas in radio galaxies (e.g. Ruffa et al. 2019a,b), consistent with a scenario in which jet-driven turbulence and heating reduce the effective inward mass flux without evacuating the gas entirely.

The decline in accretion rate propagates inward on the viscous timescale

$$t_{\rm visc} \sim \alpha^{-1}(R/H)^2\Omega^{-1},$$

which for $\alpha \sim 0.1$ and $\frac{H}{R} \sim 0.01$ corresponds to radii

$$R_{\rm fuel} \sim 10^3 - 10^4 \, R_g \sim 0.1 - 0.3 \text{ pc}$$

for black hole masses $10^8 - 10^9 \, M_{\odot}$. These radii coincide with the broad-line region and inner dusty torus, where the flow is clumpy and angular momentum transport inefficient. The observed transition timescale therefore selects sub-parsec fueling regulation. Once $\dot{m} < \dot{m}_{\rm crit}$, the inner flow rapidly becomes an ADAF on much shorter viscous timescales ($\lesssim 10^5$ yr), producing an FRII LERG while preserving the large-scale FRII morphology.

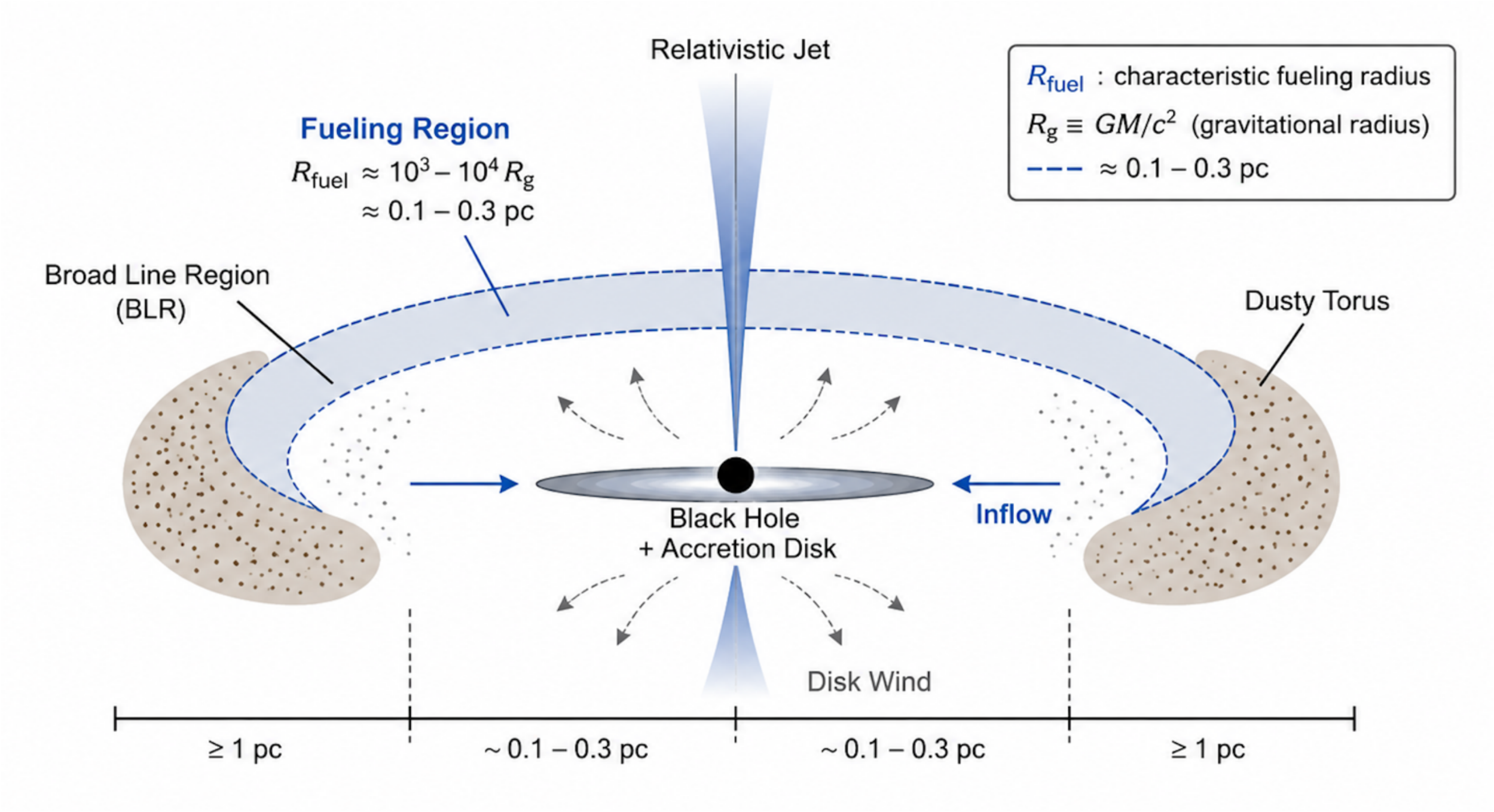

Figure 1: A central black hole is surrounded by a geometrically thin accretion disk, with relativistic jets launched along the rotation axis. The broad-line region (BLR) occupies sub-parsec scales, while the dusty torus begins at radii of order $\sim 0.1$–$0.3$ pc. The region within dashes highlights the characteristic fueling radius $R_{\rm fuel} \sim 10^3$–$10^4 R_g$, corresponding to the interface between the outer accretion flow and the BLR/torus structure. Regulation of accretion at these radii sets the viscous timescale of the flow and enables rapid transitions from radiatively efficient thin disks to advection-dominated states on Myr timescales. The Figure is AI generated.

When $\dot{m}$ at $R_{\rm fuel}$ falls below $\dot{m}_{\rm crit}$, the inner flow rapidly transitions to an ADAF on much shorter viscous timescales ($\lesssim 10^5$ yr), so the global transition time is determined by viscous propagation at the fueling radius rather than by inner-disk physics. In this framework, the FRII HERG jet heats and rarefies the ISM, reducing the effective mass supply reaching the nucleus. The resulting decline in $\dot{m}$ propagates inward through viscous evolution until the inner flow becomes radiatively inefficient, producing an FRII LERG. The key result is that the same FRII jet associated with the HERG phase also drives the transition to the LERG phase, with the required few-Myr timescale arising naturally from viscous evolution at sub-parsec radii. This extends the jet-feedback ideas proposed in Garofalo et al. (2010) and Antonuccio-Delogu & Silk (2010) by showing explicitly that they yield the required accretion-rate decline on Myr timescales.

## 3. Conclusion

In this Research Note, I examined the physical plausibility of rapid accretion-state transitions in radio-loud AGN, focusing on the evolution from FRII HERGs to FRII LERGs within the framework of Garofalo, Evans & Sambruna (2010). I showed that the required transition timescale of $\sim 4 \times 10^6$yr arises naturally if the accretion rate is regulated at sub-parsec radii, $R_{\rm fuel} \sim 0.1 - 0.3$pc ($\sim 10^3 - 10^4\ R_g$) for black hole masses $10^8 - 10^9\ M_\odot$. These scales correspond to the broad-line region and inner torus, implying that the observed transition timescale directly constrains the physical scale of the fueling process.

In this framework, FRII jet feedback heats and perturbs the surrounding medium, reducing the effective mass supply reaching the nucleus. The resulting decline in accretion rate propagates inward through viscous evolution until the flow transitions to an ADAF once $\dot{m} < \dot{m}_{\rm crit} \sim 0.01$. The few-Myr transition therefore emerges naturally from viscous evolution at sub-parsec radii rather than from gas transport on kiloparsec scales.

More generally, the framework predicts a spectrum of evolutionary timescales governed by jet power and environmental coupling. Systems with weaker feedback are expected to evolve more slowly or remain radiatively efficient for longer periods. Unlike models in which LERGs are fueled solely by large-scale hot gas accretion, the present framework naturally links radio morphology, accretion state, and cosmic evolution, providing a self-consistent explanation for the observed HERG/LERG and FRI/FRII population trends.

References

Abramowicz, M. A., Chen, X., Kato, S., Lasota, J.-P., & Regev, O. 1995, ApJL, 438, L37

Antonuccio-Delogu, V., & Silk, J. 2010, MNRAS, 405, 1303

Best, P. N., & Heckman, T. M. 2012, MNRAS, 421, 1569

Best, P. N., & Heckman, T. M. 2014, *Annual Review of Astronomy and Astrophysics*, 52, 589

Best, P. N., Ker, L. M., Simpson, C., et al. 2014, MNRAS, 445, 955

Blandford, R. D., & Znajek, R. L. 1977, MNRAS, 179, 433

Esin, A. A., McClintock, J. E., & Narayan, R. 1997, ApJ, 489, 865

Fabian, A. C. 2012, ARA&A, 50, 455

Garofalo, D., Evans, D. A., & Sambruna, R. M. 2010, MNRAS, 406, 975

Hardcastle, M. J. 2018, Nature Astronomy, 2, 273

Heckman, T. M., & Best, P. N. 2014, ARA&A, 52, 589

Hine, R. G., & Longair, M. S. 1979, MNRAS, 188, 111

Narayan, R., & Yi, I. 1994, ApJL, 428, L13

Narayan, R., & Yi, I. 1995, ApJ, 444, 231

Netzer, H. 2015, ARA&A, 53, 365

Ramos Almeida, C., & Ricci, C. 2017, Nature Astronomy, 1, 679

Ruffa et al. 2019, *MNRAS*, 484, 4239a

Ruffa et al. 2019, *MNRAS*, 489, 3739b

Shakura, N. I., & Sunyaev, R. A. 1973, A&A, 24, 337

Willott, C. J., Rawlings, S., Blundell, K. M., & Lacy, M. 2001, MNRAS, 322, 536

Yuan, F., & Narayan, R. 2014, ARA&A, 52, 529